\documentclass[prl,aps,twocolumn,groupedaddress,floats,showpacs,,nofootinbib,,preprintnumbers]{revtex4}
\usepackage{graphicx}
\usepackage{epsfig,epsf}
\usepackage{dcolumn}
\usepackage{color}
\usepackage{bm}
\newcommand{\nc}{\newcommand}
\nc{\Gz}{\fullg}
\nc{\Gvc}{\boldsymbol{G}^c}
\nc{\Gam}{\boldsymbol{\Gamma}}
\nc{\Sig}{\boldsymbol{\Sigma}}
\nc{\TS}{\tilde{\Sig}}
\nc{\TG}{\tilde{\mbox{\boldmath $G$}}}
\nc{\gam}{\boldsymbol{\gamma}}
\nc{\alp}{\boldsymbol{\alpha}}
\nc{\hs}{\hspace*{1mm}}
\nc{\fullg}{\boldsymbol{G}}
\nc{\scs}{\scriptstyle}
\nc{\beq}{\begin{eqnarray}}
\nc{\eeq}{\end{eqnarray}}
\nc{\la}{\label}
\nc{\no}{\nonumber}
\nc{\ci}{\cite}
\nc{\trace}{{\rm Tr\,}}
\nc{\setval}{\fmfset{wiggly_len}{1.5mm}\fmfset{arrow_len}{1.5mm}
\fmfset{arrow_ang}{13}\fmfset{dash_len}{1.5mm}\fmfpen{0.125mm}\fmfset{dot_size}{1thick}}

\begin{document}

%%%%%%%%%%%%%%%%%%%%%%%%%%%%%%%%%%%%%% AUTHORS %%%%%%%%%%%%%%%%%%%%%%%%%

\author{Ali Davody }

\affiliation{School of Particles and Accelerators, Institute for Research in Fundamental Sciences (IPM) \\
%P.O. Box 19395-5531
 Tehran, Iran \\
davody@ipm.ir }
%\vspace*{.4cm} {davody@ipm.ir}
%
%\vspace*{2cm}

%%%%%%%%%%%%%%%%%%%%%%%%%%%%%%%%%%%%%%%%%%%%%%%%%%%%%%%%%%%%%%%%%%%%%%%%%%%%%%
\title{Bold Diagrammatic Monte Carlo Study of $\phi^4$ Theory}
%%%%%%%%%%%%%%%%%%%%%%%%%%%%%%%%%%%%%%%%%%%%%%%%%%%%%%%%%%%%%%%%%%%%%%%%%%%%%%

\begin{abstract}
By incorporating renormalization procedure into Bold Diagrammatic Monte Carlo (BDMC), we propose
a method for studying  quantum field theories in the  strong coupling regime. BDMC  essentially samples Feynman diagrams using
local Metropolis-type updates and   does not suffer from the  sign problem.  Applying the method to  three dimensional $\phi^4$  theory, we analyze the  strong coupling limit of the theory and
confirm the existence of a nontrivial IR fixed point in agreement with prior studies.  Interestingly, we find that working with
bold correlation functions as building blocks of the Monte Carlo procedure, renders the scheme
convergent and no further resummation method is needed.
 \end{abstract}


% 02. Mathematical methods in physics

% 02.70.Ss Quantum Monte Carlo methods
% 02.70.Tt Justifications or modifications of Monte Carlo methods

% 05. Statistical physics, thermodynamics, and nonlinear dynamical
% systems (see also 02.50.-r Probability theory, stochastic
% processes, and statistics)

% 05.10.Ln Monte Carlo methods (see also 02.70.Tt, Uu in
% mathematical methods in physics; for Monte Carlo methods in plasma
% simulation, see 52.65.Pp)

\maketitle

Lattice field theory is a   well established approach for non-perturbative studies in  quantum
field theories. This  is based on the  Euclidean  path integral formulation of quantum field theory and a stochastic
 sampling of the partition function.
%Inparticular a major part of our knowledge on strong interaction comes from lattice  QCD.
This method has played a central role in developing our understanding of  strongly coupled systems including quantum chromodynamic
in particle physics and    quantum many-body systems  in  condensed matter physics.
However, the severe  sign problem is a main obstacle in applying lattice methods to systems at  finite chemical potential
or  calculating transport coefficients in the thermodynamic limit.

%prevents the use of lattice method to dealing with  systems at  finite chemical potential
%or  calculating transport coefficients in the thermodynamic limit.

%One shortcoming of lattice field theory is sign problem which occurs when we  are dealing with systems at  finite chemical
%or doing calculation in real time.

A different method based on diagrammatic formulation of field theory has been
developed  in the last few years, called diagrammatic Monte Carlo (DMC) \cite{dmc1,dmc2,dmc3}. The basic idea is
to perform a Monte Carlo process in the space of Feynman diagrams using local Metropolis-type updates.
Unlike lattice field theory, DMC samples physical quantities in the thermodynamic  limit which washes out systematic errors produced
by finite size effects. % takin continuum limit.
 However due to divergence of perturbation series, one usually needs a  resummation technique to
make the scheme convergent. This method has been applied successfully to  several systems  including polaron problem \cite{dmc2}
and the  Fermi-Hubbard model \cite{hub}.
In particular  using  the Borel resummation technique,  triviality of the  $\phi^4$ theory in four and five  dimensions as well as instability of trivial fixed point in three dimensions have  been established in \cite{Buividovich:2011zy}.

One  way of  improving  the convergence of diagrammatic Monte Carlo scheme is to expand  physical quantities in terms of
full screened (bold)  correlation functions, instead of free correlators, as is usually done   in  field theory.
This method known as Bold Diagrammatic Monte Carlo (BDMC), is shown to have a broader range of convergence \cite{bdmc1}.
 Interestingly, using BDMC, the  sign problem becomes an advantage for  convergence of the scheme.  A recent BDMC implementation
for a  strongly interacting fermionic system, namely unitary Fermi  gas, shows an excellent agreement with experimental results
on trapped  ultracold atoms \cite{nat}. In particular, the equation of state of the system at finite chemical potential has been studied,
which is hard to achieve by lattice methods due to sign problem.

In this Letter, by incorporating renormalization procedure into Bold Diagrammatic Monte Carlo (BDMC) scheme, we propose
a method for studying  relativistic quantum field theories in the  strong coupling regime.
The method is generic and applicable to any
renormalizable quantum field theory.
This  provides a new computational toolbox  for studying   longstanding problems in high energy physics, where implementing lattice
approach   suffers from the  sign problem.

 We apply the method to
  $\phi^4$ theory in three dimensions
with the following bare action

\beq
\la{Sb}
S_b =  \int d^3 x \left\{ \frac{1}{2}
\phi_{b} (x)\, (-\partial^2 + m_0^2) \, \phi_{b} (x) + \frac{g_0}{4!}
\phi^4_{b} (x)  \right\}\;,\no
\eeq

where $m_0$ and $g_0$ are bare mass and coupling respectively. It is more economical to use the
notation of \cite{Pelster:2003rc}  and rewriting the action  as follows

%For saving space, we use the following notations

\beq
\la{EF}
S_b = \frac{1}{2} \int_{12} G^{-1}_{12} \phi_1 \phi_2
+ \frac{1}{4!} \int_{1234} V_{1234}  \phi_1 \phi_2 \phi_3 \phi_4 \,,\no
\eeq

where the spatial arguments are indicated by number indices. The kernel $G^{-1}$ and potential $V$
are given by

\beq
\la{PH1}
G^{-1}_{12} &\equiv & G^{-1} ( x_1 , x_2 )  =
 \, \left( - \partial_{x_1}^2 + m^2
\right) \delta ( x_1 - x_2 ) \,\no ,\\
V_{1234} &\equiv& %V( x_1 , x_2 , x_3 , x_4 ) =
\delta ( x_1 - x_2 )\delta ( x_1 - x_3 )\delta ( x_1 - x_4 )\;\;.\no
\eeq

This model  has a nontrivial IR fixed point, first shown by Wilson and Fisher \cite{wf},
using the $\epsilon$ expansion and renormalization group techniques. According to the
RG arguments \cite{parisi,kl,zinn}, renormalized coupling, $g_r$, tends to a fixed value, $g_r^*$, when bare
coupling  becomes very large, $g_0\longrightarrow \infty $. Therefore  any non-perturbative
numerical approach to QFT, should be able to demonstrate
this feature of the theory. We show that the renormalized BDMC technique  allows us to go beyond
perturbative regime and  find   the fixed point. Interestingly, we find that working with
bold correlation functions as building blocks of Monte Carlo scheme,
renders the scheme
%makes the algorithms
convergent and no further resummation method is needed.

Our starting point is a set of Schwinger-Dyson equations for bare %one-particle irreducible
self energy, $\Sig_{b}$,  and bare one-particle irreducible four point functions, $\Gam_{b}^{(4)}$,
 derived in \cite{Pelster:2003rc}. The basic idea is  to consider a Feynman
diagram as a functional of its elements, like propagator lines. Differentiation
with respect to the free propagator, $G_b$,  leads to a  set of Schwinger-Dyson equations for correlation
functions. Using this method one finds \cite{Pelster:2003rc}

\beq
%\\[1pt]
\Sig_{b,12}  &=& - \frac{1}{2} \int_{34} V_{b,1234} \Gz_{b,34} \label{2PI}\\
&&+  \frac{1}{6} \int_{345678} V_{b,1345}  \Gz_{b,36} \Gz_{b,47} \Gz_{b,58}
 \Gam_{b,6782}\no
\\[4pt]
\Gam_{b,1234} & = & V_{b,1234}+ \tilde{\mathcal{A}}_{b,1234}+\tilde{\mathcal{B}}_{b,1234}
+\mathcal{C}_{b,1234}\;,
\label{4PI}
\eeq

in which tilde means a partial permutation on indices
\beq
\tilde{\mathcal{A}}_{b,1234}&=&\mathcal{A}_{b,1234}+\mathcal{A}_{b,1324}+\mathcal{A}_{b,1423}\;,\no\\
\tilde{\mathcal{B}}_{b,1234}&=&\mathcal{B}_{b,1234}+\mathcal{B}_{b,1324}+\mathcal{B}_{b,1423}\;,\no
\eeq

with
\beq
\mathcal{A}_{b,1234}&=&-\frac{1}{2}\,  %\int_{5678}
%\int
  V_{b,1256}  \Gz_{b,57} \Gz_{b,68} \Gam_{b,7834}\;,\no\\
\mathcal{B}_{b,1234}&=& +\frac{1}{6}\, %\int_{567890\bar{1}\bar{2}}
%\int
 V_{b,5167} \Gz_{b,69} \Gz_{b,70} \Gam_{b,902\bar{1}} \Gz_{b,\bar{1}\bar{2}}
\Gam_{b,\bar{2}348} \Gz_{b,85}\;,\no\\
\mathcal{C}_{b,1234}&=& - \frac{1}{3}\, % \int_{567890}
%\int
 V_{b,1567} \Gz_{b,58} G_{b,69} G_{b,70}   \frac{\delta \Gam_{b,8234} }
{\delta G_{b,90}}\;.\no
\eeq

%
%\beq
%\Gam_{b,1234} & = & V_{b,1234} - \frac{1}{3}  \int_{567890}
%V_{b,1567} \Gz_{b,58} G_{69} G_{70}   \frac{\delta \Gam_{b,8234} }
%{\delta G_{90}}
%-  \frac{1}{2}   \int_{5678} V_{b,1256}  \Gz_{b,57} \Gz_{b,68} \Gam_{b,7834}
%\no  \\*[3mm]
%& & - \,  \frac{1}{2} \int_{5678} V_{b,1356}  \Gz_{b,57} \Gz_{b,68} \Gam_{b,7824}
%-  \frac{1}{2} \int_{5678} V_{b,1456}  \Gz_{b,57} \Gz_{b,68} \Gam_{b,7823}\no\\
%& &+  \frac{1}{6} \int_{567890\bar{1}\bar{2}}   V_{b,5167} \Gz_{b,69} \Gz_{b,70} \Gam_{b,902\bar{1}} \Gz_{b,\bar{1}\bar{2}}
%\Gam_{b,\bar{2}348} \Gz_{b,85}
%\no  \\*[3mm]    & &
%+ \,\frac{1}{6}   \int_{567890\bar{1}\bar{2}} V_{b,5167}  \Gz_{b,69} \Gz_{b,70}
%\Gam_{b,903\bar{1}} \Gz_{b,\bar{1}\bar{2}} \Gam_{b,\bar{2}248} \Gz_{b,85} \no\\
%& & +  \frac{1}{6} \int_{567890\bar{1}\bar{2}}
%V_{b,5167}  \Gz_{b,69} \Gz_{b,70} \Gam_{b,904\bar{1}} \Gz_{b,\bar{1}\bar{2}}
%\Gam_{b,\bar{2}238} \Gz_{85}  \, .
%\label{41PI}
%\eeq

From now on   integration over repeated indices is understood.
The advantage of this set of equations is that all terms on the right hand sides  of (\ref{2PI}) and (\ref{4PI}) are  one particle
irreducible and  therefore  no irrelevant diagram  will  be produced during the Monte Carlo simulation. Also all terms are
expressed in terms of bold (exact) correlation functions except the derivative term, $\mathcal{C}_{b}$, in (\ref{4PI}). To increase the
efficiency of the method we rewrite this term using the functional chain rule and  the following identity

\beq
\label{GDE}
\fullg_{1234}^{\rm c}
= - 2 \, \frac{\delta \fullg_{12}}{\delta G_{34}^{-1}}
- \fullg_{13} \fullg_{24}- \fullg_{14} \fullg_{23} \, ,
\eeq

where $\fullg_{1234}^{\rm c}$ is the  connected four point function.
We end up with  the following \emph{bold representation} of  the derivative term

\beq
\mathcal{C}_{b,1234} & = \mathcal{D}_{b,1234}+\overline{\mathcal{D}}_{b,1234}\;,
\eeq

with

\beq
 \mathcal{D}_{b,1234}&=& - \frac{1}{3} \, %\int_{567890}
 %\int
  V_{b,1567} \Gz_{b,58} \Gz_{b,69} \Gz_{b,70}\;  \frac{\delta \Gam_{b,8234} }{\delta \Gz_{90}}\;,\no\\
 \overline{\mathcal{D}}_{b,1234} &=&  + \frac{1}{6}\,
 %\int
 V_{b,1567} \Gz_{b,58} \Gz_{b,6\bar{1}} \Gz_{b,7\bar{2}}\Gz_{b,9\bar{3}}\no\\
  & & \;\;\;\;\;\;\;\;\;\;\;\;\;\;\;\;\;\;\;\;\;\;\;\;\;\;\;\;\;\;.\Gz_{b,0\bar{4}}  \Gam_{b,\bar{1}\bar{2}\bar{3}\bar{4}}
  \; \frac{\delta \Gam_{b,8234} }{\delta \Gz_{90}}\;.\no%\\*[3mm]
\eeq

\begin{figure}%[h]
  %Requires \usepackage{graphicx}
  \includegraphics[scale=.42]{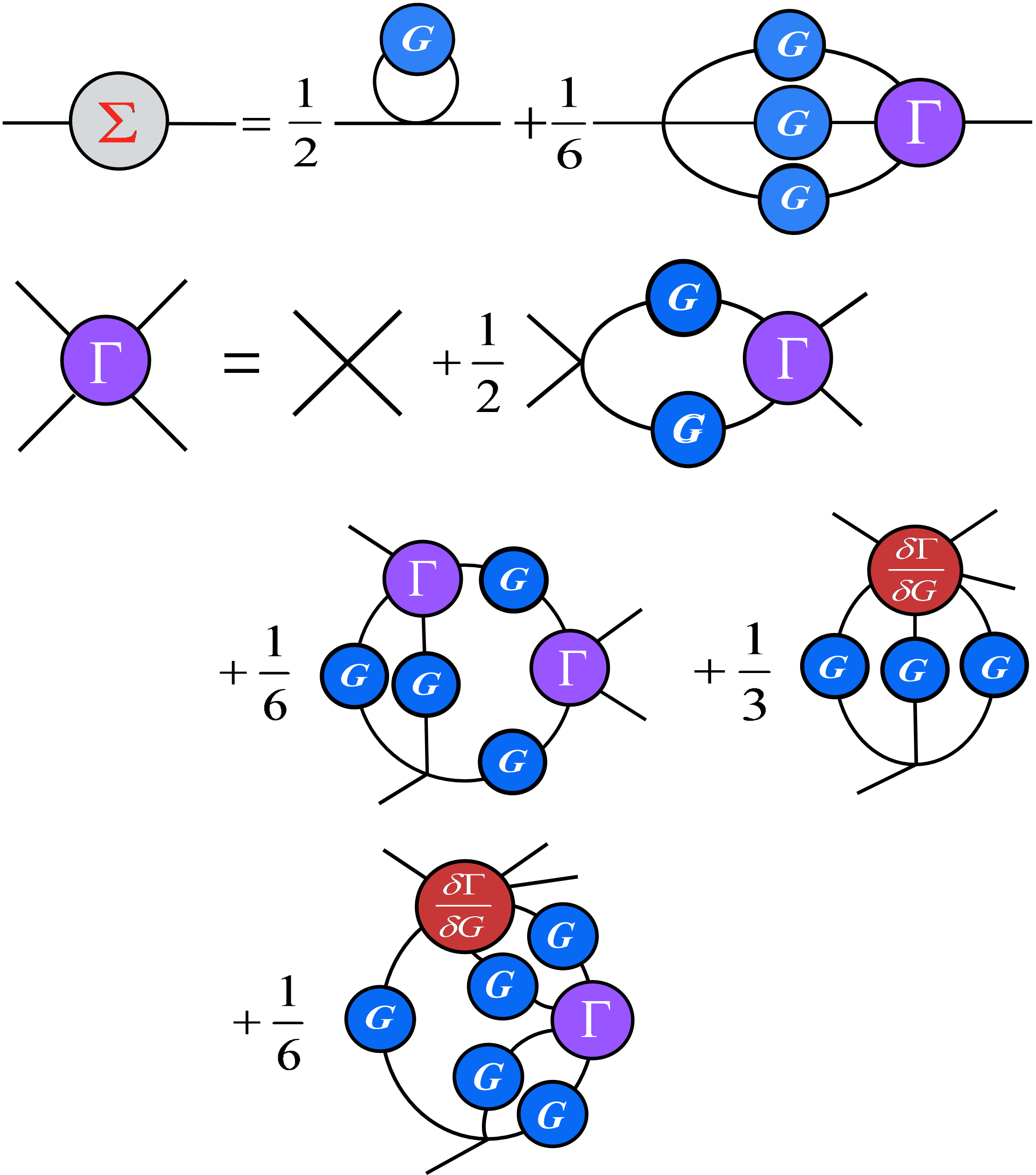}\\
  \caption{%Schwinger-Dyson equations in $\phi^4$ theory
  Bold diagrammatic expansions
  for self energy and 1PI vertex function
  in $\phi^4$ theory.
  All diagrams on the right hand sides are individually %one particle irreducible
  1PI  and expressed in terms of exact propagators.     }\label{sd}
\end{figure}

A diagrammatic  representation of (\ref{2PI}) and (\ref{4PI}) is illustrated in Fig.~\ref{sd}.

%
%\begin{figure}%[h]
%  %Requires \usepackage{graphicx}
%  \includegraphics[scale=.4]{4point.png}\\
%  \caption{}\label{sdd}
%\end{figure}

%\begin{figure}%[h]
%  %Requires \usepackage{graphicx}
%  \includegraphics[scale=.3]{final_self.png}\\
%  \caption{}\label{sddd}
%\end{figure}

By differentiating the  1PI vertex function, Eq.~(\ref{4PI}), with respect to the  full propagator, $\Gz$, we find
series expansions in terms of bold correlation functions for self energy and vertex function, in a recursive way.
In particular  for the first derivative, we have

\beq
\frac{\delta\Gam_{b,1234}}{\delta \Gz_{b,\alpha\beta}} & = &-\frac{1}{2}V_{b,12\alpha6} \Gz_{b,68}
\Gam_{b,8\beta34}\;+\alpha\longleftrightarrow \beta+\cdots\;,\no
%.
\label{41PI-moment-1deriv}
\eeq

where dots stand for  higher order  terms ( higher order in terms of the number of  bold propagators). We find   that
approximating the derivative term by the first term is sufficient for finding  the fixed point.

In order to study the behavior of the renormalized coupling constant, we translate the  Schwinger-Dyson equations
into equations  for  renormalized
correlation functions and impose renormalization conditions

\beq
\la{rc}
&\Gam_r^{(2)}\,(p^2=0)&=m^2\label{masscon}\\[3pt]
&\frac{\partial}{\partial p^2}\Gam_r^{(2)}\,(p^2)\vert_{p^2=0}&=1\label{zcon}\\[4pt]
&\Gam_r^{(4)}(0,0,0,0)&=m^{4-d}\, g_r\;. \label{grcon}
\eeq

The  renormalized proper 2-point  function which satisfies (\ref{masscon}) and (\ref{zcon}) can be written as

\beq
\Gam_{r}^{(2)}(p^2) \hs \equiv \Gz^{-1}_r(p^2)= \hs Z \,\left(\,p^2 + \mathbf{Y}(p^2) \,\right) +m^2\;,%+\frac{\partial \Sig_{r}(p^2)} {\partial p^2}|_{p^2=0} \;\;p^2 +
\label{2PI-r}
\eeq

where    $\mathbf{Y}(p^2)=\Sig_{b}(p^2=0)- \Sig_{b}(p^2)$ and  the field  renormalization constant is  given by

\beq
Z=\frac{1}{1+\frac{\partial \mathbf{Y}(p^2)} {\partial p^2}|_{p^2=0}}\;.
\eeq

Using  Eq.~(\ref{2PI}) one may rewrite $\mathbf{Y} (p^2)$ in terms of renormalized quantities as

\beq
%\begin{equation}
%\begin{split}
&&\mathbf{Y}(p^2)=\frac{g_0 Z}{6}\int\frac{d^3k}{(2 \pi)^3}\frac{d^3q}{(2 \pi)^3}\Gz_{r}(k)\Gz_{r}(q) \label{self-sc}\\
&&\times\bigg[\Gz_r(Q) \Gam_r^{(4)}(0,\overrightarrow{k},\overrightarrow{q},\overrightarrow{Q})
-\Gz_r(Q) \Gam_r^{(4)}(\overrightarrow{p},\overrightarrow{k},\overrightarrow{q},\overrightarrow{Q}) \bigg],\no
%&&\equiv \frac{g_0 Z}{6} \mathbf{\widehat{Y}}(p^2)\no
%\end{split}
%\end{equation}
\eeq

where in each term the momentum $\overrightarrow{Q}$  is determined    by the conservation of momenta that appear in the vertex function  argument.
The $\phi^4$ theory in 3 dimensions  is super-renormalizable and has only three superficially divergent diagrams, all
eliminated by the mass counter term. In addition  $\mathbf{Y}(p^2)$  is finite in any order of perturbation.
 Furthermore all vertex diagrams are superficially finite  and
we find  that it is  useful to  work with the bare form of the  schwinger-Dyson  equation (\ref{4PI}) in this case, however we have to replace bare
two-point functions with the renormalized ones.

%Note that $Y(p)$ is finite in three dimensions in any order of perturbation expansion,  and so the first term of
%self energy eq rules out in  renormalization procedure.  Also all vertex diagrams are superficially finite in three dimensions and
%we found it would be useful to  work with the bare form of schwinger-Dyson  equation in this case  (however we have to replace bare
%two-point function with the renormalized one).

Our strategy for computing the renormalized coupling constant  corresponding  to a given bare coupling is
to  solve coupled Schwinger-Dyson equations (\ref{4PI}) %, (\ref{2PI-r})
and (\ref{self-sc}) by means of general BDMC rules,
 starting with the tree level approximation for  correlation functions.  After reaching convergence,   renormalized coupling constant
 can be read off from (\ref{grcon})  by  recalling that $\Gam_r^{(4)}=Z^2\Gam_b^{(4)}$.

In order to increase
 the efficiency of the algorithm,  inspired by  the idea of worm algorithm  \cite{worm} and following \cite{svis},
 instead of sampling
the $\mathbf{Y}(p)$ and $\Gam_{b}^{(4)}$ directly, we introduce two auxiliary normalization constant terms and sample the following
quantities

\beq
I_2&=&\alpha_2+\int \mathbf{Y} (p^2)\, \Omega (p)^2 \; dp \\
I_4&=&\alpha_4+\int \Gam^{(4)}_b(p_1,p_2,p_3,\chi_{12},\chi_{13},\chi_{23})\,d\mathbf{X}
\eeq

with  $d\mathbf{X}=dp_1 \,dp_2 \,dp_3 d\chi_{12}\,d\chi_{13}\,d\chi_{23}$  where $\chi_{ij}$ is cosine  of
the angle between $\overrightarrow{p_i}$ and $\overrightarrow{p_j}$, and  $\Omega (p)$ %=\frac{c}{(p+p_0)^2}$
 is the normalized probability density that we use
to generate new momenta in Monte Carlo updates. We skip %the discussion
the    details of Monte Carlo procedure and report the results here.

%{\it Self energy}. We select new momentums $k,q $ with $\Omega (p)$ probability.

%Using the Lorentz symmetry we recast the (\ref{self-sc}) as

%\beq
%%\begin{equation}
%%\begin{split}
%\mathbf{\widehat{Y}}(p^2)&=&\int_{0}^{\infty} dk \int_{0}^{\infty}dq \int_{-1}^{+1}d\chi_1  \int_{-1}^{+1}d\chi_2 \int_{0}^{2\pi}d\widetilde{\phi}\;\Gz_{r}(k)\Gz_{r}(q) \no\\
%&\bigg[ & \Gz_r(Q) \Gam_r^{(4)}(0,\overrightarrow{k},\overrightarrow{q},\overrightarrow{Q})
%-\Gz_r(Q) \Gam_r^{(4)}(\overrightarrow{p},\overrightarrow{k},\overrightarrow{q},\overrightarrow{Q})  \bigg]\no\\
%%\end{split}
%%\end{equation}
%\eeq

%%\begin{figure}
%\beq
%\begin{fmffile}{selfenergy27}
%%\centering
%\parbox{30mm}{
%\begin{fmfgraph*}(15,8)
%  \setval
%  \fmfforce{0w,1/2h}{v1}
%  \fmfforce{3/9w,1/2h}{v2}
%  \fmfforce{6/9w,1/2h}{v3}
%  \fmfforce{1w,1/2h}{v4}
%  \fmfforce{1/2w,1h}{v5}
%  \fmfforce{1/2w,0h}{v6}
%  \fmf{plain,width=0.2mm}{v1,v2}
%  \fmf{plain,width=0.2mm}{v3,v4}
%  \fmf{double,width=0.2mm,left=1}{v5,v6,v5}
%  \fmfv{decor.size=0, label=${\scs 1}$, l.dist=1mm, l.angle=-180}{v1}
%  \fmfv{decor.size=0, label=${\scs 2}$, l.dist=1mm, l.angle=0}{v4}
%\end{fmfgraph*}}
%\end{fmffile}
%%\end{figure}
%\eeq

\begin{figure}
\vspace*{-.5cm}
\centerline{\includegraphics [scale=.4]{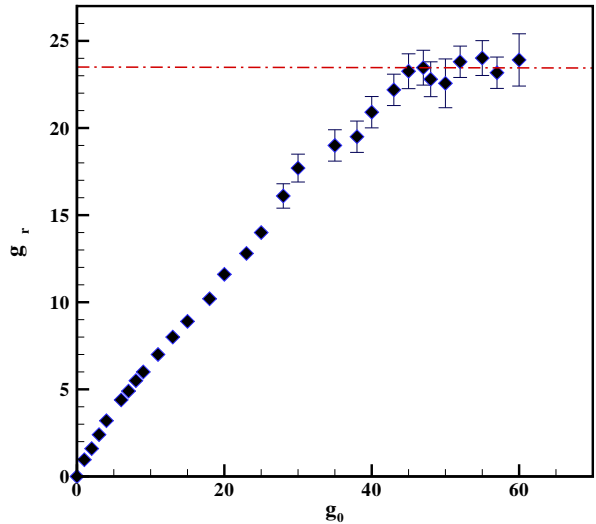}} \caption{
Result of Bold diagrammatic Monte carlo simulation
for  dimensionless renormalized coupling constant, $g_r$, as a function of bare coupling.
The asymptotic behavior of renormalized coupling is in agreement with the existence of an IR fixed point.
} \label{grplot}
\end{figure}

%\beq
%\begin{fmffile}{selfenergy56}
%\centering
%\parbox{40mm}{
%\begin{fmfgraph*}(20,18)
% \setval
%  \fmfforce{0w,0h}{v1}
%  \fmfforce{1/2w,0h}{v2}
%  \fmfforce{1w,0h}{v3}
%  \fmfforce{1/12w,1/2h}{v4}
%  \fmfforce{11/12w,1/2h}{v5}
%  \fmfforce{1/2w,1h}{v6}
%  %\fmfblob{label=$\Sigma$}{.7w}{v6}
%  \fmfv{d.sh=circle,d.f=empty,d.si=.7w,b=(1,0,1),lable=$\Sigma$}{v6}
%  \fmf{plain}{v1,v3}
%  \fmf{vanilla,width=0.2mm,left=1}{v2,v6}
%    \fmf{vanilla,width=0.2mm,right=1}{v2,v6}
%   \fmfv{decor.size=1, label=$ 1$, l.dist=1mm, l.angle=-180}{v1}
%  \fmfv{decor.size=1, label=$ 2$, l.dist=1mm, l.angle=0}{v3}
%  \fmfdot{v2}
%  %\fmflabel{$v_4$}{v4}
%\end{fmfgraph*}}
%\end{fmffile}+
%%\end{figure}
%\eeq

%\begin{figure}
%\begin{fmffile}{triangleGraph24}
%\centering
%\parbox{50mm}{
%\begin{fmfgraph*}(120,100)
%\fmfpen{thick}
%\fmfleftn{l}{2}\fmfrightn{r}{2}
%\fmfrpolyn{shaded,label=$\Gamma$}{G}{4}
%\fmfpolyn{empty,label=$K$}{K}{4}
%\fmf{fermion}{l1,G1}\fmf{fermion}{l2,G2}
%\fmf{fermion}{K1,r1}\fmf{fermion}{K2,r2}
%\fmf{fermion,left=.5,tension=.5}{G3,K3}
%\fmf{fermion,right=.5,tension=.5}{G4,K4}
%\end{fmfgraph*}}
%\end{fmffile}
%\end{figure}

Fig.~\ref{grplot} depicts the  renormalized coupling  constant as a  function of the bare coupling. As  is evident from this
plot, $g_r$ tends to an asymptotic value  in accordance with  the  renormalization group prediction.
Also the value of the fixed point, $\tilde{g}_r^*=\frac{3}{16 \pi}\,g_r^*=1.40\pm0.05$, agrees, within  error, with the
high temperature series expansion  and resummed  $\epsilon$-expansion \cite{zinn,kl}.
It is worth noticing that Fig.~\ref{grplot} provides a non-perturbative calculation of $\phi^4$ theory based on summing up Feynman diagrams.
This plot interpolates between weak and strong coupling regimes and indeed it is
not possible to produce such a result by using just perturabtive methods or RG techniques.

It is also interesting to calculate critical exponents by using  the BDMC method. Since critical exponents are related
to the scaling behavior of composite operators, we construct a new set of
Schwinger-Dyson equations   for diagrammatic expansion of composite
operators in terms of bold correlators.
For example the critical exponent $\nu$, which controls the growth  of correlation length near the phase transition, is related
to the IR behavior of the composite operator with one $\phi^2$ insertion, $\Gam^{(1,2)}$.
It is straightforward to derive  coupled equations for $\Gam^{(1,2)}$ and $\Gam^{(1,4)}$
form (\ref{2PI}) and (\ref{4PI}) by using the mass derivative trick for generating
correlation functions with $\phi^2$  insertions

\beq
\Gam^{(1,2)}_b(0, p)&=& 1-\frac{\partial}{\partial m^2} \Sig_{b}(p)\no\\
\Gam^{(1,4)}_b(0,p)&=&\frac{\partial }{\partial m^2} \Gam^{(4)}_b(p)\;.\no
\eeq

Turning on BDMC machinery it is  straightforward to solve this new set of equations in a similar way discussed for
(\ref{2PI}) and (\ref{4PI}). We postpone the numerical implementation to   future works.

%%%%%%%%%%%%%%%%%%%%%%%%%%%%%%%%%%%%%%%%%%%%
%\begin{figure}[tbp]
%%\vspace*{-1cm}
%\centerline{\includegraphics [bb=0 -20 600 870, angle=-90,
%width=1.3\columnwidth  ]{1t.ps}} \caption{(Color online).
%Scattering wave function at zero energy (solid line) and
%scattering potential (dashed line) for the attractive potential
%well with one bound state ($U_0=-3$).} \label{fig1}
%\end{figure}
%%%%%%%%%%%%%%%%%%%%%%%%%%%%%%%%%%%%%%%%%%%%
%%%%%%%%%%%%%%%%%%%%%%%%%%%%%%%%%%%%%%%%%%%%%%%%%%%%%%%%%%%%%%%%
%\begin{figure}[t]
%\vspace*{-0cm}
%\includegraphics[bb=65 380 535 430, width=\columnwidth]{fig2}
% \caption{The diagrammatic equation for the $T$-matrix (heavy dashed line) in
% terms of the vacuum $T$-matrix (light dashed line), spin-down vacuum propagator
% (straight solid line), and truncated (to the momenta {\it less} than  Fermi momentum)
% spin-up propagator (solid arc). } \label{fig2}
%\end{figure}
%%%%%%%%%%%%%%%%%%%%%%%%%%%%%%%%%%%%%%%%%%%%%%%%%%%%%%%%%%%%%%%%%%

{\it Conclusions and outlook}. In summary we described a non-perturbative simulation of a relativistic QFT, $\phi ^4$
theory in 3-dimensions, based on sampling  bold Feynman diagrams.
We used a set of coupled Schwinger-Dyson equations to expand physical quantities in terms of exact correlation functions.
The systematic method of deriving such \emph{bold expansions} in quantum field theories was proposed in \cite{klei} and used
to construct connected Feynman diagrams and to calculate their corresponding  weights in $\phi^4$ theory \cite{Pelster:2003rc} and quantum electrodynamics
\cite{Pelster:2001cc}. It is based on this fact that a complete knowledge of vacuum energy implies the knowledge of all scattering amplitudes,
`` vacuum is the world" \cite{js}.

In addition, in renormalizable QFT's it is always possible to formulate such Schwinger-Dyson equations in terms of renormalized
correlation functions and finite integrals. Combining with BDMC technique to sampling unknown functions
in terms of them, this offers a universal scheme for non-perturbative calculations in QFT's.

Applying this approach to non-Abelian gauge theories in under progress, however, one may need  more complicated
resummation methods to recover the correct physical values from truncated \emph{bold expansions}.
In the case of $\phi^4$ theory, interestingly, we observed
that without using any resummation technique, truncating the  series at lowest order, leads to convergent results.

%
%By  we found that without using any resummation
%technique, BDMC leads to convergent results.

One  way to reducing  systematic errors produced by the  truncation of bold
series is introducing a complete basis of functions and
expanding correlation functions in terms of them, $\Gz_{12}=\sum_{n,m}\; c_{n,m}\,\psi_{n,1}\,\psi_{m,2}$.
By considering $\Gam^{(4)}$  as a function of $c_{n,m}$ coefficients  the functional derivative term takes the following form

\beq
\frac{\delta \Gam_{1234}^{(4)} }{\delta \Gz_{56}}=\sum_{n,m}\,
\frac{\partial \Gam_{8234} }{\partial c_{n,m}} \;  \psi_{n,5}\,\psi_{m,6}\;.
\eeq

Performing a Monte Carlo process in space of $c_{n,m}$ coefficients  to sample the derivative term, increases  the accuracy of
the algorithm drastically.

%Since the presented  method is generic  and applicable to any renormalizable quantum
%field theory, it opens a window for studying  the longstanding problems in high energy physics, where implementing lattice
%approach is problematic  due to sign problem.

%{\it{Acknowledgment}}
The author  would  like to thank N. Abbasi, A. Fahim, A. Akhavan, D. Allahbakhsi, R.Mozafari, H. Seyedi and A.A  Varshovi for discussions. I would also like to thank  M. Alishahiha  and  A.E. Mosaffa for helpful
discussions and for carefully reading and commenting on the
manuscript.

\end{document}